\documentclass[12pt]{iopart}
\usepackage{graphicx}
\usepackage{graphics}
\usepackage{amssymb}
\usepackage{hyperref}
\usepackage{latexsym}

\def\e{\epsilon}

\def\prl{Phys. Rev. Lett.}

\def\prb{Phys. Rev. B}

\begin{document}
\title[Non affine deformations and shape recovery]{Non affine deformations and 
shape recovery in solids undergoing martensitic transformations}

\author{Jayee Bhattacharya and Surajit Sengupta}
\address{S.\ N.\ Bose National Centre for Basic Sciences, Block JD, Sector III, Salt Lake, Calcutta 700098, India}
\author{Madan Rao}
\address{
Raman Research Institute, C.V. Raman Avenue, Bangalore 560080, India, and\\
National Centre for Biological Sciences (TIFR), Bellary Road, Bangalore 560065, India}


\begin{abstract}
We study, using molecular dynamics simulations, the kinetics of shape recovery 
in a model solid undergoing transformations from square to a general rhombic 
lattice, the triangular lattice being included as a special case. 
We determine the necessary and sufficient conditions for such shape recovery
in terms of the nature and dynamics of transient and localized {\em non-affine
zones} which inevitably accompany the transformation.
\end{abstract}

Martensites\cite{marten} are classified as reversible (e.g., Nitinol) or 
irreversible (e.g., steel) according to their ability to recover their 
external shape upon thermal (or stress) cycling. What are the conditions 
under which a martensitic transformation is reversible? A recent 
argument\cite{kaunat} suggests  that a {\it necessary} condition for 
reversibility of martensitic transformations follows simply from 
{\it symmetry relations} between the parent and product phases, viz., 
reversible martensites are such that the parent and product phases are 
related by a group-subgroup relation, since then a unique parent lattice 
can be identified for every transformed product. On the other hand, 
irreversibility implies that such an identification is impossible or ambiguous.
It is clear however that the {\it necessary and sufficient} conditions for 
reversibility depend on the {\it dynamics} of transformation during thermal 
cycling. 

The conventional approach to the study of the dynamics of martensites assumes 
that the driving force for nucleation can be derived from a non-linear, 
elastic free-energy functional written in terms of a dynamical elastic strain
and its derivatives\cite{hatch}. These `strain-only' theories\cite{strn-only} 
are augmented by the condition of  local elastic compatibility, restricting  
the elastic displacements to smooth, single-valued functions. This local 
constraint automatically disallows all configurations with defects and regions 
of plasticity and assumes that the transformation is locally affine at 
{\em all} length and time scales. 
It is known, however, that there is a significant production of dislocations 
during the transformation in irreversible martensites\cite{kaunat}. 
In order to describe irreversible martensites or the transition from 
reversible to irreversible behaviour, one needs to go beyond strain-only 
theories of Ref.\cite{strn-only}.

In this paper, we use a molecular dynamics (MD) simulation of a two 
dimensional (2d) model solid \cite{ourprl,jpcm,naz} to 
(i) check the validity of the symmetry-based criterion (related to necessary 
conditions) and (ii) explore the dynamical conditions for reversibility 
(related to sufficient conditions). This is done by simply changing a single 
potential parameter which allows us to explore {\it both} reversible and 
irreversible martensites within the same model.

Our 2d model solid\cite{ourprl,jpcm,naz} is composed
of $N$ particles interacting via a repulsive, {\em anisotropic}
2-body potential $(V_2)$, parametrized by an anisotropy coefficient $\alpha$,
and a short-range 3-body potential $(V_3)$,  of strength $v_3$,
which favours local square configurations. Specifically,
$V_2({\bf r}_{ij};\alpha) = v_{2}\left(
\sigma_0/r_{ij}\right)^{12} \lbrace 1 + \alpha \cos^{2} 2\theta_{ij}\rbrace
$, where ${\bf r}_{ij}$ is the relative displacement between particles
$i$ and $j$, $\theta_{ij}$ the angle between ${\bf r}_{ij}$ and an
arbitrary external axis, and $V_3({\bf r}_{ij},{\bf r}_{jk}; v_3) =v_{3}
\left[f_{ij} f_{jk} \sin^{2}4\theta_{ijk} + {\rm permutations}\right] $, with
$\theta_{ijk} = \cos^{-1}\{{\bf r}_{ij}\cdot{\bf r}_{jk}/(r_{ij}r_{jk})\}$.
Energy and length scales are set using
$v_2 = 1$ and $\sigma_0 = 1$ and the unit of time is $\sigma_0\sqrt{m/v_2}$,
where $m$ is the particle mass, which for typical values, translates to a
MD time unit of $1 ps$. Decreasing $V_3$ induces a square (Sqr) to 
rhombic (Rmb) transition while $\alpha$ controls the apex angle of the 
Rmb phase from $\pi/6$ at $\alpha =0$, i.e. a triangular (Trg) lattice 
to $\pi/2$ for $\alpha > 1.2$.

The equilibrium and dynamical properties of structural transitions for this 
model in the constant number - $N$, area $A$ and temperature $T$ ensemble 
with fixed external shape have been studied in some detail in 
\cite{ourprl,jpcm,naz}. We briefly mention the main conclusions below in order 
to set the stage:
\begin{enumerate} 
\item The transformation from the Sqr to Rmb (or Trg) phase progresses 
by heterogeneous nucleation and growth.

\item The transformation, or {\em order parameter} (OP),  strain 
$e_T$ (shear for Sqr $\to$ Rmb) also produces a {\em non-order parameter} 
(NOP) volumetric strain $e_V$ which is slaved to $e_T$. The NOP strain 
introduces non-local interactions between the transformed regions leading to 
the formation of a twinned microstructure typical of martensites for low 
transformation temperatures.
 
\item Localized {\em non-affine zones}
or NAZs are generated at the transformation front where the displacement
of the atoms cannot be described using  affine deformations (viz.
scaling and shear). In this model solid the NAZs are in the NOP sector. 

\item The NAZs in the NOP sector are created when the volumetric stress 
$\sigma_V$ exceed a threshold. As soon as the NAZs form, $\sigma_V$ tends to 
decrease thereby screening the non-local interactions. At low temperatures, 
the dynamics of the NAZs in the frame of the moving front is slow so that this 
screening is never total for martensites. At high temperatures, on the other 
hand, the NAZs quickly cover the entire growing nucleus destroying the twinned 
pattern leading to an untwinned, disordered and irreversible {\em ferrite}. 

\item Particle trajectories within the NAZs tend to be ordered when 
martensites are obtained while they are disordered for the high temperature 
ferrite.
\end{enumerate}     

In this study we are interested exclusively in shape transformations 
accompanying the formation of martensite so the temperature is set to a low 
value ($T = 0.1$) throughout. Further we need an isolated solid with stress 
free boundaries to allow for deformations of the shape of the solid as it 
transforms. We associate every particle with a ``glue'' density, $\xi(r) = 1$ 
for $r \leq r_g$ dropping smoothly to zero at $ r = r_g$ (chosen to be the 
next-nearest neighbour distance). The embedding energy of particle $i$ in this 
glue is $V_g = -K_g \, \sum_{j=1,N} \xi({\bf r}_{ij})$, where $K_g$ is the 
cohesive energy. The glue causes the particles to stick to each other producing
a 2d solid whose boundary is self consistently determined by the many-body 
interaction among particles alone\cite{note}. The value of $K_g$ is tuned so 
as to maintain the density to be roughly constant across the transformation
-- the area $A$, on the other hand, may vary at fixed pressure and $T$. 
\begin{figure}
\begin{center}
\includegraphics[width=8.0cm]{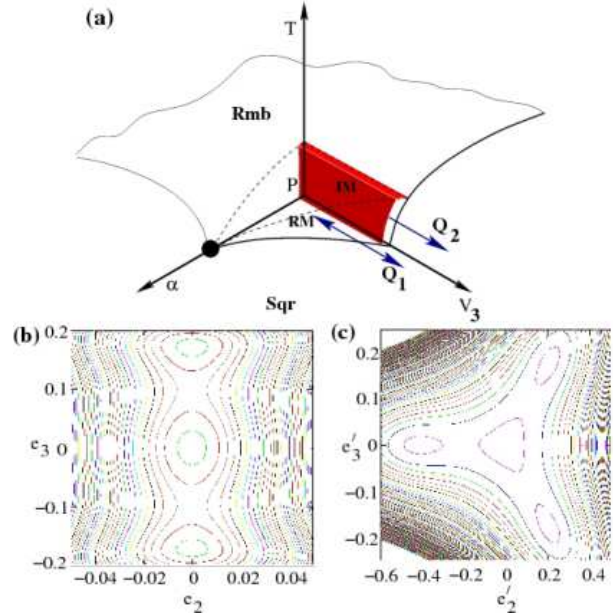}
\end{center}
\caption{
Schematic dynamical phase diagram in $T-v_3-\alpha$, showing equilibrium phases
(square (Sqr) and rhombus (Rmb)) by solid lines. The triangular (Trg) solid 
exists for $\alpha = v_3 = 0$ -- point $P$. The black dot marks the 
location of a tricritical point at which the jump in the order parameter 
vanishes. The dynamical phase martensite is formed for quenches at $T$ below 
the curved surface bounded by the dashed lines. Dynamical martensitic phases 
upon cycling: reversible (RM) and irreversible (IM) martensite (red region)
are shown. The quench and cycling protocols ${\bf Q_1}$ and ${\bf Q_2}$ are 
denoted by arrows. (b) Zero temprature energy per particle $E/N$ for 
$\rho = N/V = 1.05$, $\alpha = 0$ and $v_3 = .408$ 
as a function of the OP strains $e_2$ and $e_3$ near the Sqr $\to$ Rmb 
transition showing a metastable Sqr minimum at (0,0) and two degenerate, 
stable Rmb minima. (c) $E/N(e_2^{\prime},e_3^{\prime})$ 
for $\rho = N/V = 1.05$, $\alpha = 0$ and $v_3 = 3.034$ for the reverse 
transformation from the Trg to Sqr phase. The strains are now calculated 
from the Trg phase at $v_3 = \alpha = 0$.      
}
\label{ebyn}
\end{figure}
Our MD simulation\cite{ums} uses
a leap-frog Verlet algorithm with a time step of $.001$ which conserves
energy to 1 in $10^{-6}$, and a Nos\'e-Hoover thermostat to obtain
trajectories of particles in the constant N, area and T ensemble. This gives
us an equilibrium phase diagram in $T-v_3-\alpha$\cite{jpcm,naz,jayee}, which 
we display schematically in Fig. 1(a). We start with an equilibrium square 
solid at $t=0$ and cycle the control parameters as shown by the arrows (
${\bf Q_1}$ and ${\bf Q_2}$ in Fig. 1(a)).

The variation of the $T = 0$ energy per particle $E(e_2,e_3)/N$ as a
function of the OP strains $e_T = (e_2,e_3)$ where $e_2$ is the deviatoric and 
$e_3$ the shear strain is
is shown in Fig. \ref{ebyn}(b) and (c). For a general first order
{\em p4m} $\to$ {\em p2} transition, there are four minima apart
from the minimum at $(0,0)$ corresponding to the Sqr phase. For the
special case of Sqr $\to$ Rmb transition shown in Fig. \ref{ebyn}(b), 
$e_2 = 0$ and the four minima collapse into two.
Note that the product Rmb phases are connected in OP
space to an unique parent Sqr phase so that the reverse transformation
is also unique. If $\alpha = 0$, (Fig.\ref{ebyn}(c)) however, 
decreasing $v_3$ finally leads us to the Trg structure which has a 
{\em higher} symmetry ({\em p6m}) than Sqr, there being three symmetry axes.
Measuring the strains from the Trg lattice, we find that the strain energy now 
has four minima. The central one at (0,0) corresponds to the
Trg lattice which is surrounded by the {\em three} degenerate Sqr minima
only {\em one} of them being the original parent Sqr phase.

In our model solid we can change the group-subgroup relation of the 
parent-product by changing $\alpha$.
\begin{figure}
\begin{center}
\includegraphics[width=9.0cm]{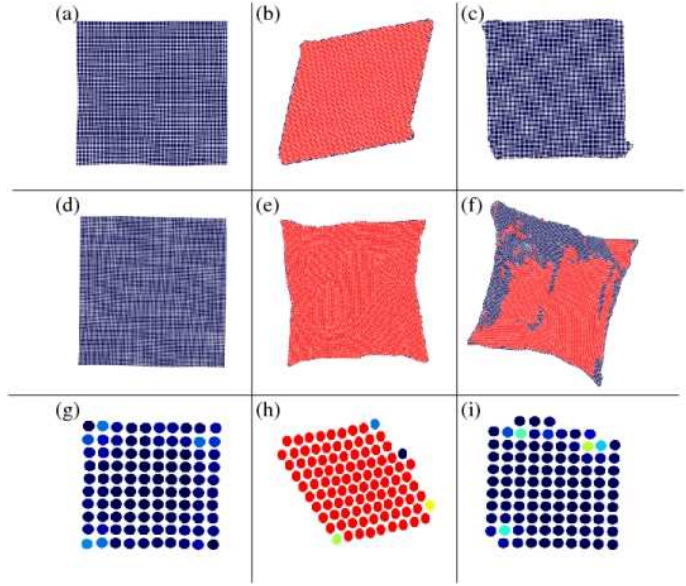}
\end{center}
\caption{(color-online)
Configurations of particles as $v_3$ is first reduced and then increased 
taking the system from the Sqr to the Rmb phase and back.
The colors correspond to the quantity 
$\omega_i = \omega_0 \sum_{jk} \sin^2 (4 \theta_{ijk})$ 
where the particles $j$ and $k$ are near neighbors of $i$ and $\omega_0$
is chosen so that $\omega$ varies from $1$ (red) in the Rmb/Trg to $0$ (blue) 
in the Sqr phase. (a)-(c) ${\bf Q_1}$: Configurations at $T=0.1$,$\rho=1.05$,
$\alpha=1.25$,$N=14400$.(a) $v_3=10$ (b) $v_3 = 0.5$ (c) $v_3 = 10$.
(d)-(f) ${\bf Q_2}$: same as above with $\alpha = 0$ with $v_3 = 10$ (d), 
$v_3 = 0.5$ (e) and $v_3 = 6$ (f). (g)-(i) same as (d)-(f) except 
$N = 100$, the small size of the system makes the transformation reversible.
}
\label{mart}
\end{figure}
Keeping $\alpha = 1.25$ we decrease $v_3$ to $v_3 = 0.5$, in steps 
of $.5$ holding the system for $10^4$ MD steps at each $v_3$ 
(${\bf Q_1}$ in Fig.\ref{ebyn}(a)). The Sqr $\to$ Rmb phase
transition at $v_3^{\ast} \approx 1.$ breaks the (Ising) symmetry
between the two degenerate Rmb minima. 
Since the presence of the surface breaks translational symmetry nucleation
predominately proceeds from the surface or the corners.
The crystal structure as well as the overall shape of the crystallite 
transform from Sqr to Rmb (Fig.\ref{mart} (a)-(c)). In the reverse path when 
$v_3$ is increased again, the shape change reverses in accord with 
Ref.\cite{kaunat} to Sqr. The situation, on the other hand, is quite different 
when $\alpha = 0$ shown by the line ${\bf Q_2}$ in Fig.\ref{ebyn}(a). The 
Sqr $\to$ Rmb phase transition occurs at $v_3^{\ast} = 1.4$ and as 
$v_3$ is further reduced to zero, finally a Trg lattice results. The overall 
shape of the product crystal is not rhombic (Fig.\ref{mart}(d)-(f)), because 
the Ising symmetry is not completely broken and grain boundaries exists 
between different degenerate variants of the triangular phase forming at 
different portions of the crystallite, -- the grain boundary energy between 
the variants of the Trg product being low. During the reverse transformation, 
these grains tend to transform to different Sqr lattices {\em not necessarily 
the one which produced them in the first place}, setting up large internal 
stresses, which are accomodated initially by shape deformations but
cause the solid to rupture\cite{tin} when excessive.

\begin{figure}
\begin{center}
\includegraphics[width=12.0cm]{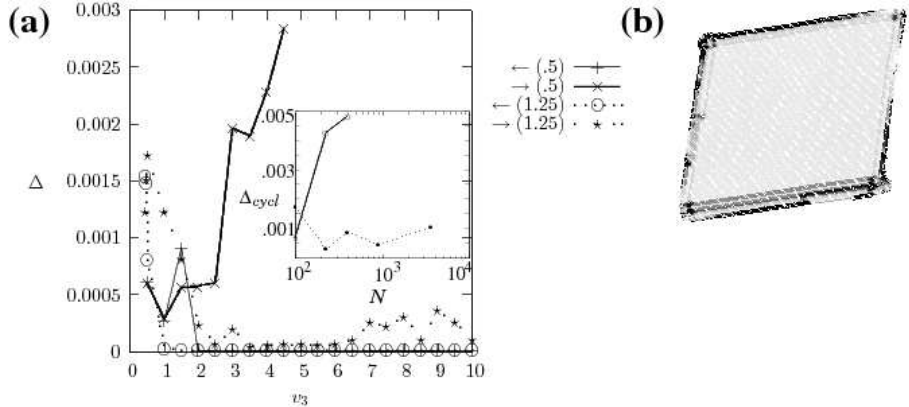}
\end{center}
\vskip -.5cm
\caption{
(a) Overlap parameter $\Delta$ vs. $v_3$ at $T=0.1$,$N=14400$, and $\rho=1.05$
for various values of $\alpha$ plotted during the forward ($\leftarrow$, decreasing $v_3$)
and reverse ($\rightarrow$) paths. The meaning of the symbols used are
explained in the key to the left. (inset)
Shape recovery as measured by $\Delta_{cycl}$ -- the difference in values of 
$\Delta$ after one cycle as a function of the
size of the solid $N$ for $\alpha = .5$
$K_g = 1$ (full line) and $K_g = 2$ (dashed line).
Note that large cohesive energy ($K_g$) helps shape recovery. (b) NAZs for
the Rmb $\to$ Sqr reverse transformation for $\alpha = 1.25$. Darker regions 
have larger $D_{\Omega}^2$. Note that the NAZs are confined mainly near the 
surface.
}
\label{ovr}
\end{figure}
The overall shape change of the sample during a quench cycle, may be quantified 
as follows. We first identify boundary atoms by counting the number of nearest 
neighbors. The coordinates of the selected atoms are then used to define the 
shape function $r_{S}(\theta)$ with the angle $\theta$ measured from the 
$x-$ axis. Denoting $r_S^{(0)}(\theta)$ as the shape of the initial square 
configuration, we have the overlap $\Delta$ given by,
\begin{equation}
\Delta = \frac{1}{16L^2}\int d\theta [ r_S(\theta) - r_S^{(0)}(\theta)]^2
\end{equation}
where $L$ is the system size and we have taken care to check for configurations
related to the initial configuration by global translations and rotations. The 
result is plotted in Fig.\ref{ovr}(a) for two 
values of $\alpha$ as $v_3$ is first reduced from a large initial value
($= 10$) to $0.5$ and then increased again for
a system of $N = 14400$ particles. As expected, we observe that for
$\alpha =1.25$,  the transformation is perfectly reversible. For $\alpha = .5$ 
when the product resembles the
Trg phase, $\Delta$ tends to increase during the reverse transformation
showing that the shape change becomes irreversible. The value of $\Delta$
at the end of one cycle $\Delta_{cycl}$ measures overall shape recovery 
at the end of the cycle. In the inset of Fig.\ref{ovr}(a) we plot 
$\Delta_{cycl}$ for a system of particles at $\alpha = .5$ as a function of 
$N$. In view of the conclusions drawn from Ref.\cite{kaunat}, it is striking 
that the shape transformation, while irreversible for large $N$, becomes 
reversible as $N$ decreases. Increasing the cohesive 
energy $K_g$ from $1$ to $2$ also reduces $\Delta_{cycl}$ and
aids shape recovery.

We now demonstrate that all of these observations may be understood from 
the dynamics of NAZs which inevitably accompany the solid - solid structural 
transformation\cite{naz}. As in Ref.\cite{naz} we study the dynamics of the 
NAZs by computing the local non-affine parameter  using the 
following procedure. For every particle $0$ in 
the Trg configuration, we define a neighbourhood $\Omega$, using a  
cutoff equal to the range of the potential ($\sim 2.5 \sigma_0$). This is 
compared with that of the same particle in the transformed lattice by defining 
the parameter\cite{langer}, 
\begin{eqnarray}
D^2_{\Omega}({\bf r},t) & = & \sum_{i \in \Omega} \sum_m [ r_i^m(t) - r_0^m(t) - \sum_n (\delta_{mn} + \e_{mn}) \nonumber \\
        &  &  \times (r^n_i(0) - r^n_0(0)) ]^2 
\end{eqnarray}
which needs to be minimized with respect to choices of affine strains 
$\e_{mn}$. Note that the indices $m\,{\rm and}\,n = 1,2$ and $r^n_i(0)$ and 
$r^n_i(t)$ are the $n^{th}$ component of the position vector of the 
$i^{th}$ particle in the reference (Sqr) and transformed lattice, respectively.
The {\em residual} value of $D^2_{\Omega}({\bf r},t)$ is a measure of 
{\em non-affineness}. 

\begin{figure}
\begin{center}
\includegraphics[width=12.0cm]{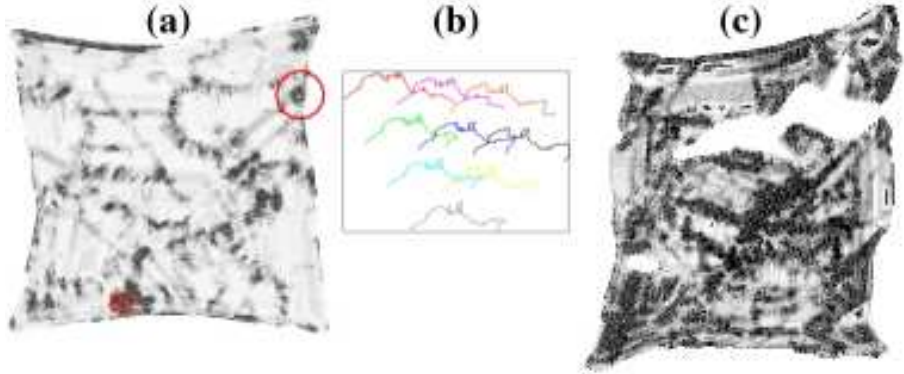}
\end{center}
\caption{
(a)The non affine parameter $D^2_{\Omega}$ (see text) for a Trg lattice during 
the reverse transformation at $v_3 = 5$, $\alpha = 0$, $\rho = N/L^2 = 1.05$ 
and $T = 0.1$. Dark regions correspond to large $D^2_{\Omega}$ within 
non-affine zones (NAZ)s which surround isolated and disjoint regions where 
$D^2_{\Omega}$ is small. (b) Particle trajectories in a NAZ -- red square in 
(a). Note that individual trajectories are disordered. (c) The same system as 
in (a) at a later time and $v_3 = 7.5$. Note that the system fractures along 
lines with high $D^2_{\Omega}$, the crack nucleates at a spot on the surface,
shown by the red circle in (a), where $D^2_{\Omega}$ is particularly large.   
}
\label{chi}
\end{figure}

The transformation connecting the Sqr and Rmb phases is accompanied by NAZs as 
shown in Fig.\ref{ovr}(b). Firstly, as in \cite{ourprl} and \cite{naz} these 
NAZs are in the NOP sector, being associated with the local volume strain 
$e_V$. They are mainly localized near the surface and rapidly disappear 
as the crystal transforms from Sqr to the Rmb phase, being advected 
out at the completion of the transformation. Secondly, as in Refs.\cite{ourprl}
and \cite{naz},  particles close to the NAZs within the transformed region 
move ballistically and in a coordinated manner. It is these two properties of 
the NAZs discussed here that ultimately renders the Sqr $\to$ Rmb martensitic 
transformation reversible, in spite of significant transient and localized 
plastic deformation.
 
Even the slightest amount of plasticity in the OP sector, on the other hand, 
would make the transformation irreversible. Within our model system, a deep 
quench to the $\alpha = 0$, $v_3 = 0$ region 
produces a Trg solid which is not related to the parent square 
lattice by a group-subgroup relation\cite{hatch}. During the reverse 
transformation, therefore, there is no unique parent lattice that the system 
can revert to. This produces non-affineness in the OP sector due to a 
multiplicity of affine paths and destroys reversibility. 
The result from our MD simulations is shown in 
Fig.\ref{chi}(a). It is interesting to note that the largest values of $D^2$ 
exists along boundaries of isolated patches within which the transformation is 
to a single square lattice. Particles at the boundary of these patches are 
{\em structurally frustrated} and tend to follow separate possible affine paths
along trajectories which are disordered and ``diffusive'' {\em despite the 
temperature being low} (Fig.\ref{chi}(b)). Unlike in reversible martensites,
the creation of these NAZs, {\em do not} reduce stress. Indeed, the stress 
continues to increase being set by the OP strain and eventually the crystal 
fractures (Fig.\ref{chi}(b)). Examination of the fracture surface shows that 
the solid breaks apart precisely along regions of large $D_{\Omega}^2$ so that 
the NAZs provide seeds for the heterogenous nucleation of cracks.   
In essence, therefore, microstructural reversibility in
martensites, is related to the nature of the accompanying plastic deformation.

Is is now easy to see why shape recovery is restored for smaller and stiffer 
solids. Since NAZs are produced when the local stress increases beyond 
a threshold, the average distance between NAZs is determined by the statistics
of the stress threshold and of the local stress. In general one expects that 
NAZs are separated by some average lengthscale $l_c$ determined by the 
yield strength of the material, with stiffer materials having larger $l_c$. 
It may thus be possible to avoid OP strain induced NAZs completely either by 
reducing the system size or increasing $l_c$ by making the material 
stronger.     

In this paper, we have examined the problem of shape recovery of a solid 
undergoing a structural transition from a Sqr to either Rmb ($p4m \to p2$) or 
Trg ($p4m \to p6m$) lattice. We obtain necessary and sufficent conditions 
for shape recovery by examining local regions of plasticity viz. NAZs 
produced during transformation. In agreement with Ref.\cite{kaunat} we show 
that a group - subgroup transformation is always reversible since it is 
accompanied by reversible NAZs related to the slaved NOP strains. For group - 
nonsubgoup transformations, in general, shape change is not recoverable due to 
the presence of plasticity in the OP sector, {\em unless} the size of the 
system is smaller than the typical distance between NAZs. We believe that 
our work has relevance to applications of real martensites\cite{marten}
as well as on the general question of phase ordering dynamics of solid 
state transformations.      
 
{\bf Acknowledgements\,} Discussions with A. K. Raychaudhuri, and 
K. Bhattacharya are gratefully acknowledged. We thank the Unit for Nanoscience 
and Technology, SNBNCBS and the Department of Science and Technology, 
Govt. of India for financial support.


\vskip .5 cm
  

\begin{thebibliography}{99}
\bibitem{marten}
A.\ Roitburd, in {\it Solid State Physics}, ed.\ Seitz and Turnbull
(Academic Press, NY, 1958)\,; 
{\em Martensite} eds. G.\ B.\ Olson and W.\ S.\ Owen, (ASM International,
The Materials Information Society, 1992).

\bibitem{kaunat}
K. Bhattacharya, S. Conti, G. Zanzotto and J. Zimmer, Nature, {\bf 428},
55 (2004).


\bibitem{hatch}
D. M. Hatch, T. Lookman, A. Saxena, and H. T. Stokes, \prb,{\bf 64}, 060104(R),
(2001).

\bibitem{strn-only}G. R. Barsch {\em et al.}, \prl {\bf 59}, 1251 
(1987); K. \O.  Rasmussen {\em et al.}, \prl {\bf 87},
055704 (2001); T. Lookman {\em et al.}, \prb 67, 024114 (2003), and
references therein.

\bibitem{ourprl}
M. \ Rao and S. Sengupta, \prl {\bf 91}, 045502 (2003).

\bibitem{jpcm}
M. Rao and S. Sengupta, J. Phys: Condens. Mat. {\bf 16},
7733 (2004).

\bibitem{naz}
J. Bhattacharya, A. Paul. S. Sengupta and M. Rao, arXiv:0706.3321v2  

\bibitem{jayee}
J. Bhattacharya, S. Sengupta and M. Rao, preprint


\bibitem{ums}
D. Frenkel and B. Smit, {\it Understanding Molecular Simulations}, 2$^{\rm nd}$
Edition, (Academic Press, California, 2002)

\bibitem{note}The contribution of the glue potential to the total energy is
$K_g \sum_{i} \rho_i \xi_i = K_g \sum_i \rho_i \sum_j K_{ij} \rho_j$, where $K_{ij}=1$ for nearest neighbour $i$ and $j$ and zero otherwise. This reduces to $K_g \sum_{i} \rho_{i} - {\cal {O}}(1/L)$, where the boundary contributions scale 
down with system size $L$. In the thermodynamic limit, this simply leads to a 
resetting of the chemical potential.

\bibitem{drop}
M.\ Rao and S.\ Sengupta, \prl {\bf 78}, 2168 (1997);
M. Rao, and S. Sengupta,  Curr. Sc. {\bf 77}, 382-387 (1999);
S. Sengupta and M. Rao, Physica (Amsterdam) {\bf 318A}, 251 (2003).

\bibitem{tin} Disintegration of real materials due to internal transformation
stresses is well known. See for example, http://en.wikipedia.org/wiki/Tin

\bibitem{langer}
M. L. Falk and J. S. Langer, Phys. Rev. E {\bf 57}, 7192 (1998)

\end{thebibliography}
\end{document}